# MECHANICAL STABILITY STUDY FOR INTEGRABLE OPTICS TEST ACCELERATOR AT FERMILAB*

M.W. McGee†, R. Andrews, K. Carlson, J. Leibfritz, L. Nobrega, A. Valishev
Fermi National Accelerator Laboratory, Batavia, IL 60510, USA

## Abstract

The Integrable Optics Test Accelerator (IOTA) is proposed for operation at Fermilab. The goal of IOTA is to create practical nonlinear accelerator focusing systems with a large frequency spread and stable particle motion. The IOTA is a 40 m circumference, 150 MeV (e-), 2.5 MeV (p+) diagnostic test ring. A heavy low frequency steel floor girder is proposed as the primary tier for IOTA device component support. Two design lengths; (8) 3.96 m and (2) 3.1 m long girders with identical cross section completely encompass the ring. This study focuses on the 3.96 m length girder and the development of a working prototype. Hydrostatic Level Sensor (HLS), temperature, metrology and fast motion measurements characterize the anticipated mechanical stability of the IOTA ring.

## INTRODUCTION

The Integrable Optics Test Accelerator (IOTA) given in Figure 1 is a small storage ring (40 m circumference) being built at Fermilab [1–3]. Its main purposes are the practical implementation of nonlinear integrable lattices in a real machine, the study of space-charge compensation in rings, and a demonstration of optical stochastic cooling.

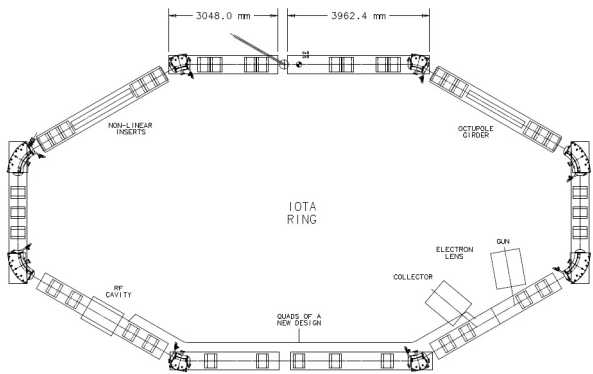

Figure 1: Layout of IOTA Ring.

The concept of nonlinear integrable optics applied to accelerators involves a small number of special nonlinear focusing elements added to the lattice of a conventional machine in order to generate large tune spreads while preserving dynamic aperture [4]. The concept may have a profound impact in the design of high-intensity machines by providing improved stability to perturbations and mitigation of collective instabilities through Landau damping [5]. Table 1 provides a summary of IOTA parameters.

Table 1: Summary of IOTA Ring parameters.

| PARAMETER | VALUE |
| --- | --- |
| Nominal P+ Beam Kinetic Energy | 2.5 MeV |
| Nominal P+ Beam Intensity | $8 \times 10^{10}$ |
| Transverse P+ Emittance (rms) | 1-2 μm |
| Nominal E- Beam Energy | 150 MeV |
| Nominal E- Beam Intensity | $1 \times 10^{9}$ |
| Transverse E- Emittance (rms) | 0.1 μm |
| Circumference | 40 m |
| Bending Field | 0.7 T |
| Beam Pipe Aperture | ϕ50 mm |

As the mechanical design for the IOTA Ring matures, magnet components, girders and adjustable stands are being procured and fabricated. A prototype of the floor girder was developed early in this process for evaluation. Two concepts were considered for basic (foundational) support; concrete base and rigid steel construction. The steel construction concept was chosen due to overall cost, ease of implementation and increased available space. Also, studies involving the enclosure's environment in terms of stability were considered. Both slow and fast motion studies began once the prototype girder was installed.

## IOTA RING DESIGN

The girder is a structural steel design which repurposes supporting stands acquired from MIT-Bates. These supporting structures that are made from (2) 524 mm [20-5/8"] wide x 825.5 mm [32-1/2"] length x 25.4 mm [1"] thick plates, a 304.8 mm [12"] x 304.8 mm [12"] x 12.7 mm [½"] wall x 838.2 mm [33"] long square tube, (4) 12.7 mm [½"] thick 45 degree gussets applied top and bottom. Two different girder lengths are considered: 3.96 m [156"] and 3.1 m [120"] long design. Each MIT-Bates stand was cut 508 mm [20"] from the bottom and (2) parallel 9.5 mm [3/8"] wall thickness x 203 mm [8"] x 254 mm [10"] rectangular structural tube lengths are welded together along each girder length, top and bottom at the center.

The basic design from the ground up considers a floor variation of +/- ¼" and the application of Quikrete® Non-Shrink Precision Grout. Actual grouting will follow the component installation and alignment. Each MIT-Bates stand was initially attached to the floor using (4) M19 diameter [¾"-10] Hilti rod and a drop-in. Above the Hilti rod, a leveling nut with washer will set the elevation (nominally 31.25 mm [1-1/4"] from the floor) of each MIT-Bates stand and a compensating nut and washer attachment exists on top of the plate with a slight applied torque to secure.

## GIRDER STABILITY

In general, limiting motion of components is not a priority of the support system design as relative motion is of greatest concern. Thermal and motion studies involved instrumenting the IOTA enclosure with temperature sensors on the walls, and applying a HLS system. Independent HLS systems were positioned along adjacent enclosure walls; North-South (18 m span), East-West (12 m span) and on top of the prototype floor girder (3 m span). The prototype girder will also be applied as a support system for the IOTA electron lens during initial development [6]. Table 2 provides a summary of IOTA error sources.

Table 2: Summary of error sources [7].

| PARAMETER | VALUE |
|---|---|
| Beta Functions | 0.01 (relative) |
| Phase Advance | 0.001 |
| BPMs | 100 μm |
| Bunch Transverse Size | ~100 μm |
| Bunch Length | 2 cm |
| FFT (ΔQ) | ~1 x $10^4$ |
| Energy loss during 1 ms sampling window | --- |
| Errors in NL potential | --- |
| Overall | δA ~0.1 mm, δQ~1 x $10^4$ |

### IOTA Enclosure Study

Air velocity measurements were conducted within the IOTA enclosure in order to quantify thermal-convective effects present. The enclosure air temperature and velocity profiles will change as the IOTA Ring is commissioned and brought into operation. The tunnel leading up to the IOTA enclosure will be populated as the Fermilab Accelerator Science and Technology (FAST) beamline develops, restricting air flow. Also, operating Injection and IOTA devices, although water cooled, will raise the average temperature by 5 °C and may cause fluctuations of 2 °C.

An E+E, EE220 humidity and temperature transmitter with accuracy ±0.1 °C and operating range of -40 to 80 °C was used within the IOTA enclosure. Air velocity varied vertically as the greatest velocities were found between 1 and 3 m from the floor. The nominal girder height was 0.73 m (where velocities were lower) and subsequently, the air flow had less of an impact regarding thermal stability of each girder. There were stagnate air sections of the IOTA ring, therefore a FEA of the thermal differences was completed, since relative motion is important. Figure 3 depicts a 3.96 m long girder supporting (2) 60° dipole magnets at the ends (worst case loading), elevated from 20 °C to 25 °C. A temperature difference of 1° C will cause a longitudinal thermal expansion or contraction of 48 μm, over the 3.96 m girder length.

Enclosure temporal air temperature measurements were completed using thermometry with the IOTA enclosure. The IOTA enclosure maintained a temperature of 19 ± 0.5 °C. The vertical temperature-gradient fluctuations are negligible, however air velocities had a tendency to vary where flows were above 0.2 m/s. ANSYS Workbench was used to calculate temperature profile and corresponding directional displacement due to thermal strain as shown in Figure 2.

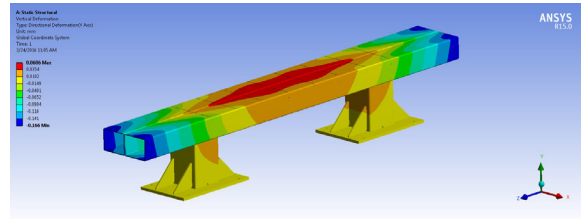

Figure 2: ANSYS result for thermal difference study.

### Thermal Inertia

Thermal Inertia (or mass) was studied through heat transfer analytical estimates and FEA. Worst case convective conditions measured during the enclosure study were used in the models. FEA results confirmed that for an enclosure air temperature change of ±0.5 °C with 1 hour time cycle, a floor girder experiences only $1/5^{th}$ of the ambient temperature change.

## SLOW & FAST MOTION MEASUREMENT

Fermilab has implemented several types of HLS systems in various locations over the past fifteen years. These devices utilize connected reservoirs of distilled water in order to determine the relative difference in elevation (device location to location). These HLS systems were developed through a collaboration between Fermilab and Budker Institute [8]. IOTA enclosure fast motion measurements were completed using tri-axial blocks of HS-1 geophones and vertical Sercel L-4c seismometers, as shown in Figure 3.

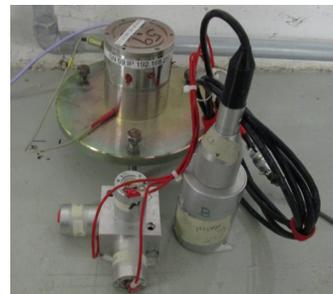

Figure 3: HLS, tri-axial geophone and seismometer.

The SAS-E type HLS device measure the capacitance of the reservoir in order to track the relative height of the water over time. This SAS-E HLS device utilizes a Power-over-Ethernet switch and TCP/IP protocol to provide a readout for each sensor which is data-logged through Accelerator Controls Network (ACNet) control system. The HLS systems used are capable of accuracies of 1 μm or greater, which is not influenced by geometric distance (if certain requirements are met, such as closed system and free surface) [10]. Typical data are shown in Figure 4. The data show MINOS floor tilt, the tidal motion with a 12.6 hour period is visible in the data.

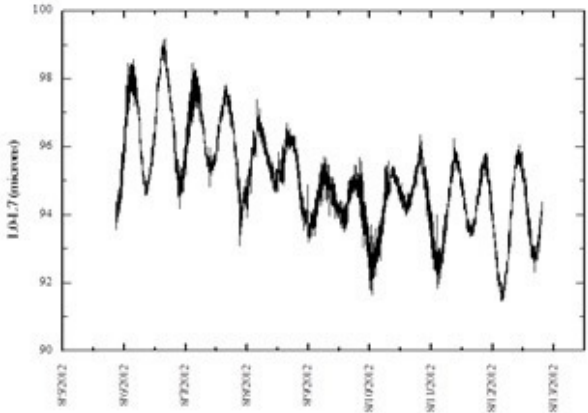

Figure 4: Typical MINOS HLS data [9].

The vertical integrated rms displacement shown in Figure 5 indicates that the enclosure motion was ~ 0.1 mm and also that only a slight amplification at the floor girder level (762 mm above) occurred. The transfer function between the floor and girder was 1.2.

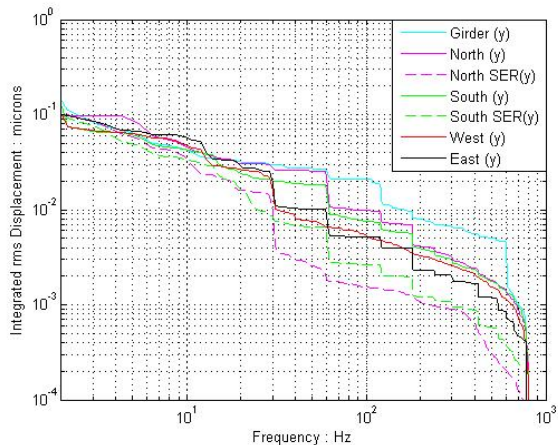

Figure 5: Enclosure vertical integrated rms displacement.

## FUTURE WORK

Dipole magnet (primary component) placements have been superimposed onto the floor as monuments and beamline center for the IOTA Ring has been established. Installation of the floor girder system is on-going as the girder fabrication nears completion.

## ACKNOWLEDGEMENTS

We wish to thank the SRF Technical Group; Chris Exline, Wayne Johnson, Ron Kellet, Elias Lopez and Craig Rogers for their valuable assistance. Also, thanks to Greg Bulat and Gary Markiewicz regarding machine and weld shop support and Designers Fred Mach and Steve Wesseln. Also, thanks to the Fermilab Alignment and Metrology Group (AMG) staff: Virgil Bocean, Gary Crutcher, Mike O'Boyle, Gary Teafoe and Chuck Wilson.